\documentclass[twocolumn,conference]{IEEEtran}
\usepackage[T1]{fontenc}
\usepackage[latin9]{inputenc}
\usepackage{url}
\usepackage{graphicx}
\usepackage{multirow}

\usepackage{amsmath}
\usepackage{soul}
\begin{document}

\newcommand{\scalecircuit}{0.31}
\newcommand{\scalevisio}{0.75}
\newcommand{\scaleplot}{0.49}

\setlength{\textfloatsep}{10pt}
\setlength{\topskip}{10pt}

\sloppy

\title{A Pulse Width Modulation based Power-elastic and Robust Mixed-signal Perceptron Design}

\author{
    Sergey Mileiko\textsuperscript{\dag{}},
    Rishad Shafik\textsuperscript{\dag{}},
    Alex Yakovlev\textsuperscript{\dag{}},
    Jonathan Edwards\textsuperscript{\ddag{}}\\
    \textsuperscript{\dag{}}Newcastle University, UK;
    \textsuperscript{\ddag{}}Temporal Computing, UK
}

\maketitle

\begin{abstract}
Neural networks are exerting burgeoning influence in emerging artificial intelligence applications at the micro-edge, such as sensing systems and image processing. As many of these systems are typically self-powered, their circuits are expected to be resilient and efficient in the presence of continuous power variations caused by the harvesters.
In this paper, we propose a novel mixed-signal (i.e. analogue\slash digital) approach of designing a power-elastic perceptron using the principle of pulse width modulation (PWM). Fundamental to the design are a number of parallel inverters that transcode the input-weight pairs based on the principle of PWM duty cycle. Since PWM-based inverters are typically agnostic to amplitude and frequency variations, the perceptron shows a high degree of power elasticity and robustness under these variations. We show extensive design analysis in Cadence Analog Design Environment tool using a $3$ x $3$ perceptron circuit as a case study to demonstrate the resilience in the presence of parameric variations.
\end{abstract}


\section{Introduction and Motivation}
 \label{sec:introduction}


Perceptron is the basic building block of deep neural networks used in machine learning applications~\cite{Hagan:1997:NND:249049}~\cite{366063}~\cite{doi:10.1111/j.1467-8667.1989.tb00026.x}. It consists of an input vector, a set of weights and a bias to produce binary classification outcomes, as follows:
\begin{equation}
f(x) = \begin{cases}
      1, & \text{if}\ \textbf{w}.\textbf{x} + b > 0 \\
      0, & \text{otherwise}
    \end{cases}
\end{equation}
where $w$ is a vector of real-valued weights, \textbf{w}.\textbf{x}  is the dot product $\sum_{i=1}^{m}w_i x_i $ with $m$ number of inputs, and $b$ is the bias. The process of deciding the appropriate weights (\textbf{w}), often also known as training, serves as the basic principle of supervised learning. When $m$ becomes large, it approximates the behaviour of a biological neuron.

\begin{figure}[ht]
    \centering
    \includegraphics[scale=\scalevisio]{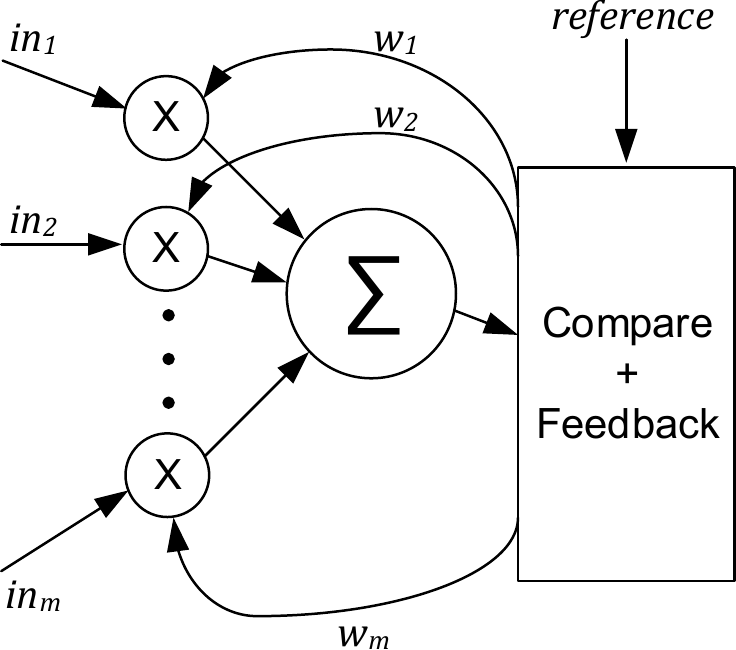}
    \caption{The structure of perceptron.
}
    \label{fig:perc_struct}
\end{figure}

Fig.~\ref{fig:perc_struct} shows the typical structure of a perceptron~\cite{HUNG19913}~\cite{330393}. The core of it is an adder that adds $m$ weighted inputs. The result of the addition is compared with a reference during the training phase. During this time, the weights are updated to ensure the reference is matched
.

For hardware implementation multiplication and addition are crucial arithmetic circuits in a perceptron
~\cite{1243906}. Such arithmetic operations require significant area and power costs, which depend on the number of input-weight pairs and the precision of the multipliers and adders.

Future micro-edge applications will be increasingly autonomous. The key for autonomy is being not only based on smartness through machine learning capability but also being able to work from energy-harvesting sources. In other words, the circuits must be capable of working under a dynamic range of power variation.

For a class of applications involving machine learning at the micro-edge, such as sensors with data filtering and compression, we can envisage power to be extracted from the environment~\cite{8330023}. Energy-harvesting power sources may not be always equipped with accurate power regulation circuits, which are themselves power-hungry. Hence, in this work we are aiming at developing a perceptron design that is resilient to power variations, i.e. power-elastic~\cite{benafa-async2018}.

Existing perceptron designs are predominantly digital; although a number of analogue implementations have been reported~\cite{7551398}~\cite{isscc_2016_chen_eyeriss}. Nonetheless, these designs are vulnerable to power supply variations. As such, these are not suitable for working under extreme power variations. In other words, these have poor power elasticity properties that refrain them from providing useful computation under unreliable power supplies.


To address power elasticity, which gives reliable results even with unstable supply voltage and input frequency we propose to transfer the arithmetical computation process from the digital domain to the temporal domain, where the information is encoded in the input pulse width. This would guarantee the reliability of the input data, because the pulse width (input signal duty cycle) is not affected neither by the input frequency, nor by its amplitude.

The aim of this paper is to design a power and frequency elastic perceptron, which performs the arithmetic computation in the PWM-coded format. The main \textbf{\textit{contributions}} are:
\begin{enumerate}
    \item a mixed-signal perceptron design using duty cycle based temporal weight encoding and input switching via a PWM inverter, and
    \item extensive validation experiments in Cadence Analog Design tool demonstrating its resilience in the presence of amplitude and frequency variations. 
\end{enumerate}

The rest of the paper is organized as follows. Section II presents the proposed design approach. Section III validates the approach using a number of parametric sweeps to demonstrate power elasticity and resilience. Finally, Section IV discusses and concludes the paper. 

\section{Proposed Approach}
\label{sec:approach}

The proposed approach is based on the principle that if the input of an inverter is a periodic signal, such as clock, the average voltage on its output is inversely proportional to the duty cycle of the input clock. This is due to the fact that during the interval of time when the input is Low the output capacitance is charged with current from the power source via the PMOS transistor, and during the interval of input being High the capacitance is discharged via the NMOS transistor.
Since an inverter is a digital component, whose output equals to logic $'0'$ or $'1'$, it should be "analogized" (i.e. transcoded) in order to convert the input duty cycle into the output voltage that is a corresponding proportion of the supply voltage. This may be achieved by the following ways:

\begin{itemize}

\item increasing the input switching frequency,

\item increasing the output capacitance, and

\item limiting the output current.

\end{itemize}

\begin{figure}[ht]
    \centering
    \includegraphics[scale=\scalevisio]{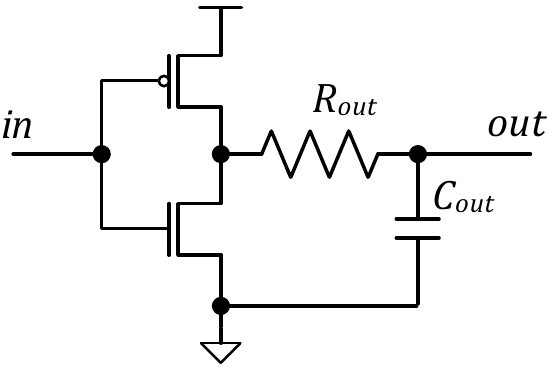}
    \caption{An inverter with the output resistor and capacitor.
}
    \label{fig:inv_res}
\end{figure}

Fig.~\ref{fig:inv_res} shows the inverter circuit that meets the requirements. The output capacitance of the inverter has been increased by adding a capacitor $C_{out}$ between the output of the inverter and ground. The output resistor $R_{out}$ performs several functions. Firstly, it limits the current, increasing the capacitor's charging\slash discharging time. Secondly, it reduces the system's power consumption (See Section III). And, lastly, it adds linearity to the output characteristics as the PMOS and NMOS resistance may be different with different drain voltages. A large resistive load can neglect this difference, the demonstration of which will be shown in Section III.

A key feature of this circuit is that if we connect the outputs of several cells, the resulting output voltage will be inversely proportional to the average value of the inputs duty cycle. Therefore, using these inverters, we can build an adder with the PWM-coded inputs, leading to analog output.

\begin{figure}[ht]
    \centering
    \includegraphics[scale=\scalevisio]{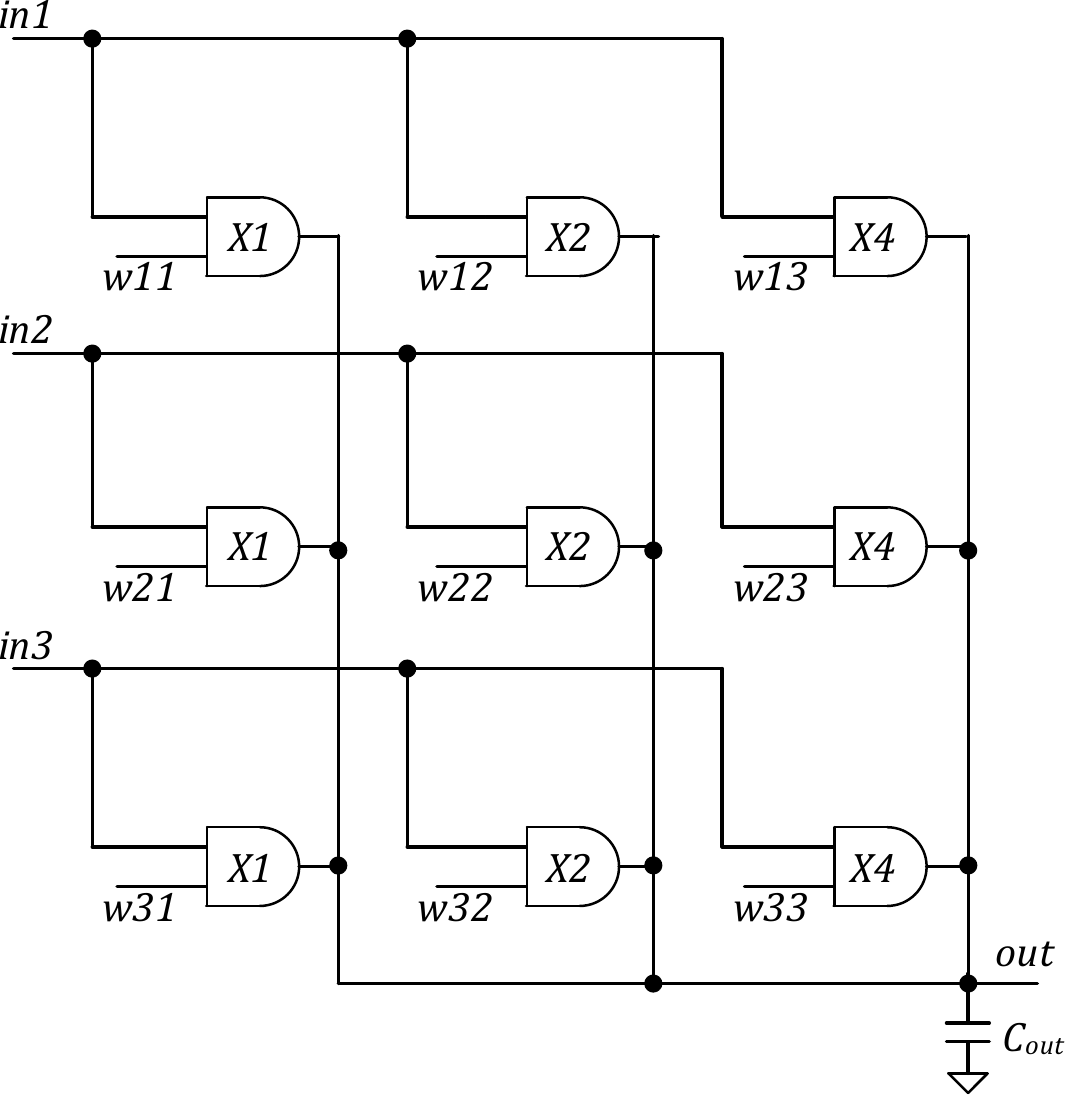}
    \caption{3x3 weighted adder.
}
    \label{fig:3x3}
\end{figure}

To design a perceptron the ability to integrate weighted adders is another crucial design requirement. The adders must be capable of programming the input weights, when required. This is performed by replacing the inverters by AND gates. One input of this gate is the PWM-coded, and another is a digital switch for enabling or disabling this cell. Fig.~\ref{fig:3x3} shows a perceptron architecture with $3\times 3$ weighted adder, built with such gates.

As can be seen, the circuit adds 3 PWM-coded inputs multiplied by 3-bit weights. Every weight bit is implemented on a separate cell. The least significant bit goes to the cell with the smallest transistor size and the largest output resistor~(cells $X1$). The second bit is computed at the cell with doubled transistors width and halved output resistance~(cells $X2$). And the most significant bit is coded with $4$ times wider transistors, and $4$ times smaller output resistor~(cells $X4$).

The output voltage of this adder is calculated as follows:

\begin{equation}
\label{eq:output_formula}
	V_{out} = (V_{dd} - GND) \cdot \frac{\sum_{i=1}^{k} DC_{i} \cdot W_{i}}{k \cdot (2^{n}-1)}.
\end{equation}
where $k$ is the number of the inputs, $n$ is the number of bits of the weight, $DC_i$ is the duty cycle of the input~$i$, and  $W_i$ is the weight of the input~$i$.

The transcoding of spatial data (in digital form) to temporal domain (in PWM duty cycle), and the mixed-mode (analogue\slash digital) multiplier\slash adder operation have significant impact on the circuit complexity, and its resilience in the presence of amplitude and frequency variations. These will be extensively validated in the next section.

\section{Experimental Results}
\label{sec:results}

A prototype circuit (based on Fig. 2 and 3) is designed using UMC65nm technology and simulated in the Cadence Analog Design Environment tool.

The ability of an inverter to convert a PWM-coded signal into analog is demonstrated by the following experiment. The circuit from Fig.~\ref{fig:inv_res} has been simulated with the parameters listed in Table I. These parameters have been optimized after extensive sweep experiments. For brevity, these optimization experiments are not reported here.

\begin{table}
\centering
\caption{Simulation parameters used in experiments}
\begin{tabular}{|c|c|}
\hline
 Input signal frequency & $V_{dd}=2.5V$\\\hline
 Transistors width & $n_{width}=320nm$, $p_{width}=865nm$\\\hline
 Transistors length & $n_{length}=p_{length}=1.2um$\\\hline
 Output capacitor & $C_{out}=1pF$\\\hline
\end{tabular}
\end{table}

\begin{figure}[ht]
    \centering
    \includegraphics[scale=\scaleplot]{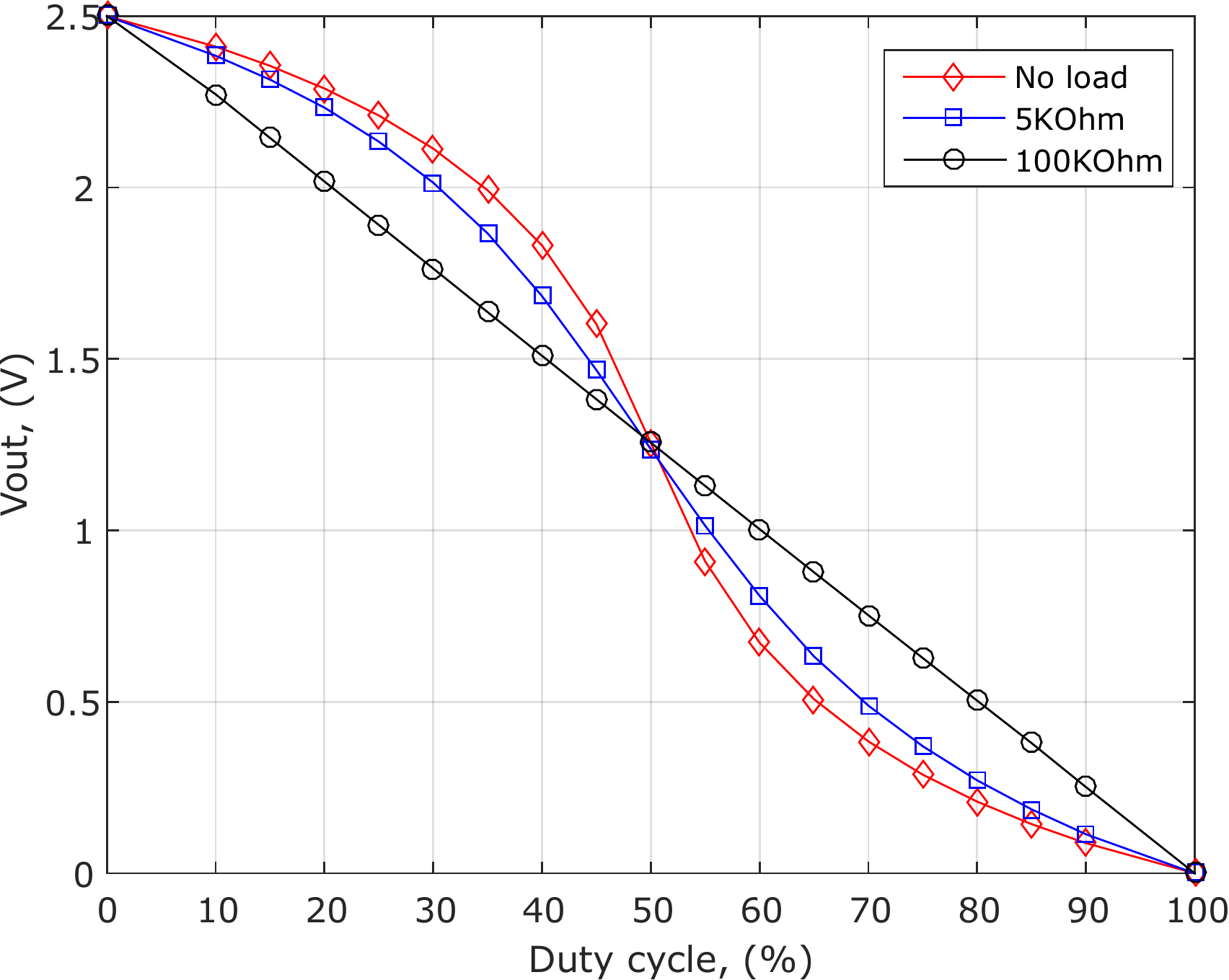}
    \caption{Output voltage vs input duty cycle of the inverter cell.
}
    \label{fig:out_DC}
\end{figure}

Fig.~\ref{fig:out_DC} shows the dependency of the output voltage from the input signal duty cycle for different sizes of the output load resistor $R_{out}$.

The plot shows that the output voltage is reversely proportional to the input signal duty cycle. However, this proportionality is not linear for the inverter with small load resistor and without it. This is caused by the non-linearity of the transistors. Their resistance depends on their drain-to-source voltages, and can be different with different $V_{out}$. In the case of the large output resistor, it brings the greatest contribution to the overall resistance, and the output function becomes purely linear.

\begin{figure}[ht]
    \centering
    \includegraphics[scale=\scaleplot]{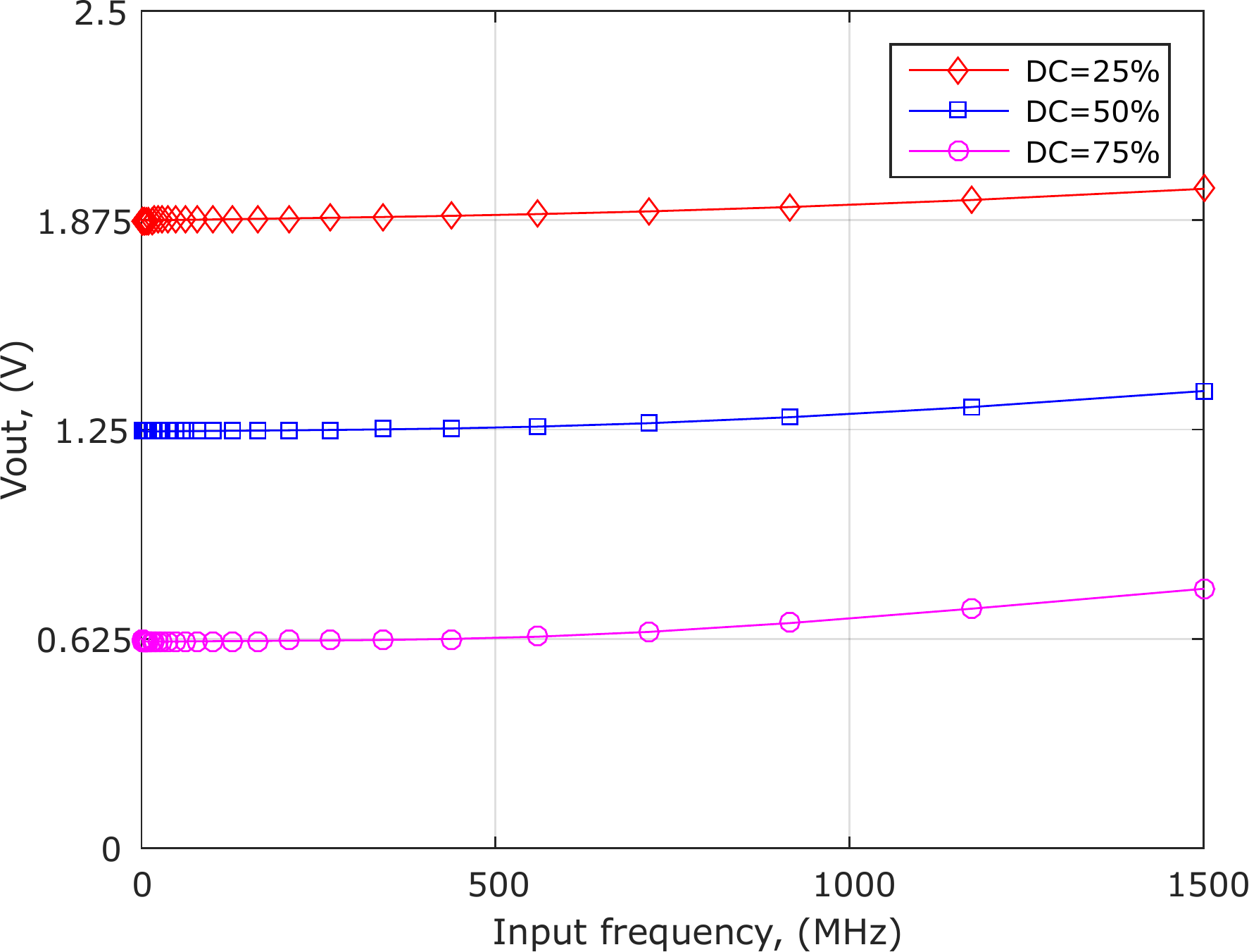}
    \caption{Output voltage vs input frequency of the inverter cell.
}
    \label{fig:out_freq}
\end{figure}

Fig.~\ref{fig:out_freq} demonstrates the resilience of the cell to the input frequency variation. Parameters of the simulation are the same an in the previous one (see Table I). The circuit uses fixed value of the output resistor $R_{out}=100KOhm$. The plot shows the output voltage for the input frequencies from 1MHz to 1500MHz, and input duty cycles 25\%, 50\%, and 75\%. As can be seen, the values of $V_{out}$ are almost the same for a wide range of frequencies. This demonstrates a high degree of frequency resilience for the proposed perceptron design.

\begin{figure}[hb]
    \centering
    \includegraphics[scale=\scaleplot]{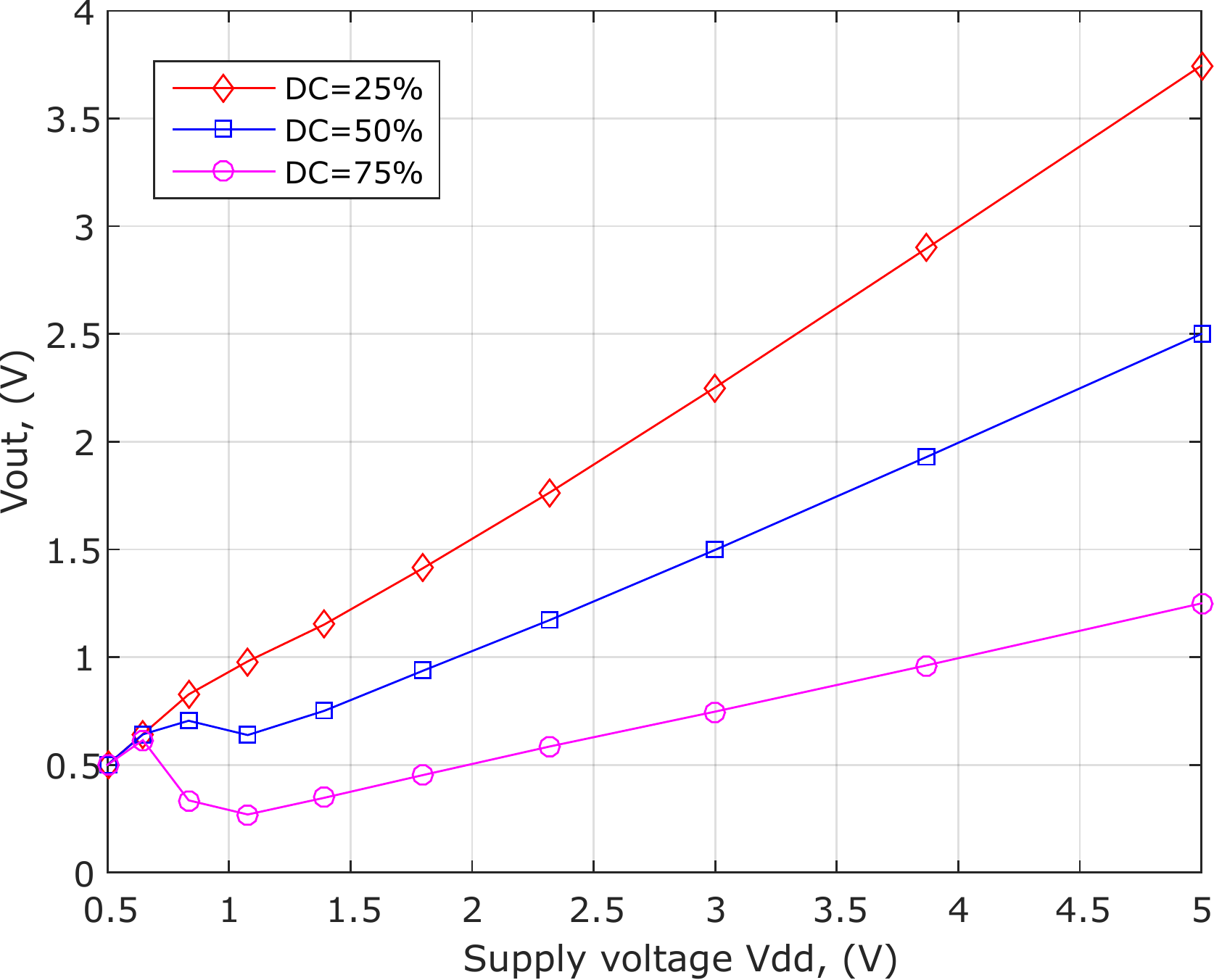}
    \caption{Output voltage (absolute value) vs power supply.
}
    \label{fig:out_vdd_abs}
\end{figure}

To demonstrate the perceptron resilience to the power variations we simulated the inverter circuit with different values of the supply voltage and input amplitude. The results are shown in Fig.~\ref{fig:out_vdd_abs}. The simulations parameters are the same as shown in Table I. The input signal frequency $f_{in}$ is constant and equals to 500MHz.

\begin{figure}[ht]
    \centering
    \includegraphics[scale=\scaleplot]{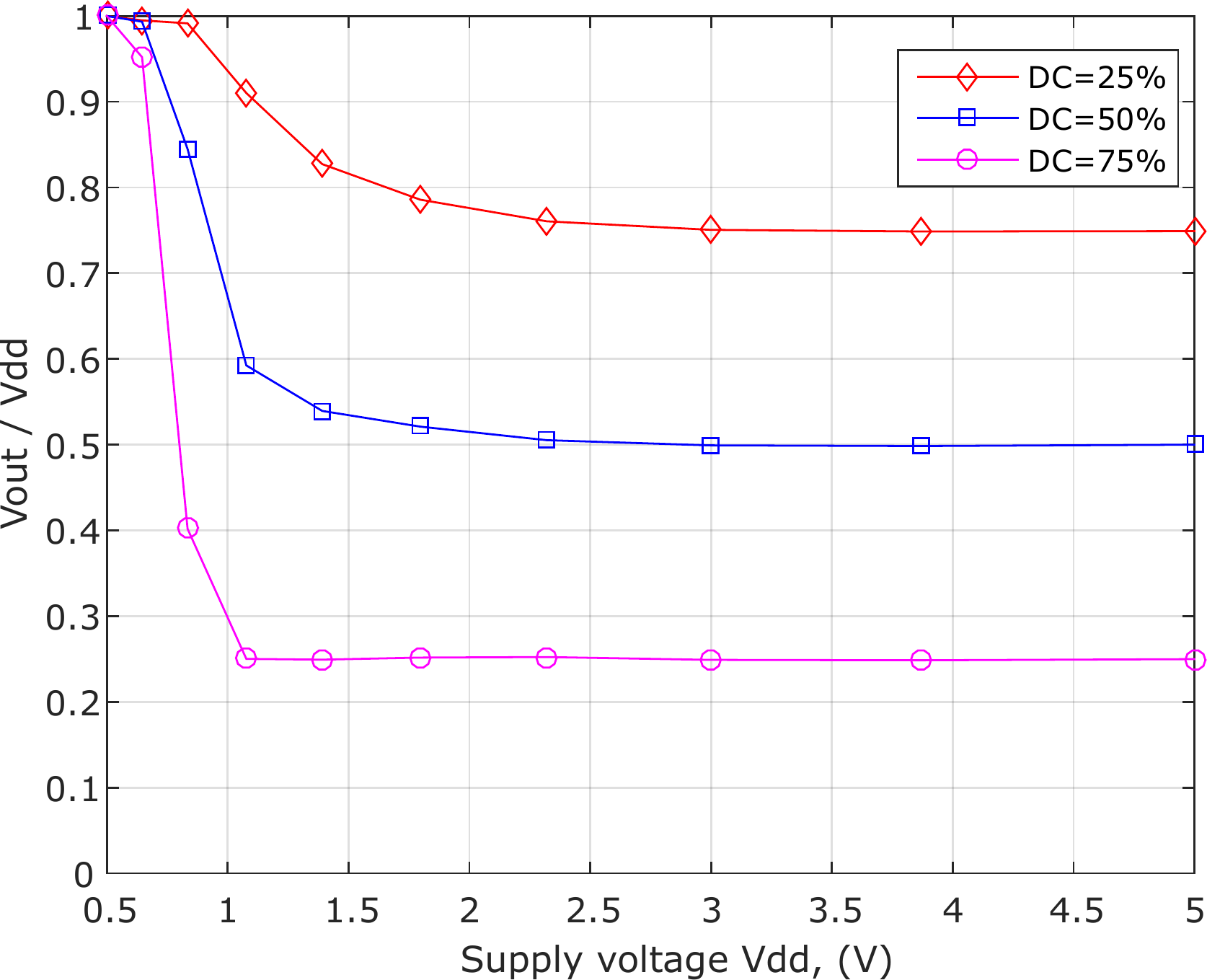}
    \caption{Output voltage (relative to the power supply) vs power supply.
}
    \label{fig:out_vdd}
\end{figure}

As can be seen, the output voltage grows almost linearly with increased $V_{dd}$. As expected, higher duty cycle shows lower output voltage trends, or vice versa. In the case of the unstable supply voltage, the absolute value of the output voltage does not bear any reliable information. In this case, we should consider the relation between the output voltage and the supply voltage. This relation will be proportional to the input duty cycle independently from the $V_{dd}$. This is demonstrated by Fig.~\ref{fig:out_vdd} where the $y$ axis represents not the absolute value of $V_{out}$, but the relation of $V_{out}$ to $V_{dd}$ that is more relevant for unstable power conditions. 

The circuit demonstrates high resilience to the supply voltage variations. Starting from 1 - 1.5V the relationship of the $V_{out}$ to $V_{dd}$ remains the same for different duty cycles of the input signal.

The simulations below demonstrate the correctness of operation of the $3\times 3$ weighted adder shown in Fig.~\ref{fig:3x3}. The parameters of the simulations are the same as in previous, except the size of the output capacitor that has been extended to 10pF. In this simulation we set up different values of three inputs (their weight and duty cycle), and compared the resulted output voltage with its theoretical values (calculated using the formula~\ref{eq:output_formula}). The results are shown in Table~\ref{tab:3x3_simulations}. 

\begin{table}[h]
    \centering
    \caption{The results of the $3 \times 3$ weighted adder.}
    \label{tab:3x3_simulations}
    \begin{tabular}{| c | c | c | c | c | c | c | c | }
        \hline
        \multirow {2}{*}{DC1}         & \multirow {2}{*}{W1} & \multirow {2}{*}{DC2} & \multirow {2}{*}{W2} & \multirow {2}{*}{DC3} & \multirow {2}{*}{W3} & $V_{out}$  & $V_{out}$ \\
         & & & & & & theoretical &  simulation \\
        \hline
         70\% & 7 & 80\% & 7 & 90\% & 7 & 2.00V & 1.99V \\
        \hline
        50\% & 1 & 50\% & 2 & 50\% & 4 & 0.42V & 0.39V \\
        \hline
        20\% & 5 & 60\% & 6 & 80\% & 7 & 1.21V & 1.17V \\
        \hline
        95\% & 7 & 90\% & 6 & 80\% & 6 & 2.00V & 2.05V \\
        \hline
        30\% & 1 & 40\% & 4 & 50\% & 2 & 0.34V & 0.29V \\
        \hline
        80\% & 7 & 20\% & 3 & 50\% & 4 & 0.96V & 0.89V \\
        \hline
    \end{tabular}
\end{table}

The simulations results correspond to the theoretical ones, however, the relative error is quite large, especially for the lower output voltages. Despite of this, such errors are still affordable, especially, in the case of perceptron that is a-priori not accurate.

The simulations have been conducted with various input frequencies in the range from $1MHz$ to $1GHz$, but the frequencies did not have any effect on the results, and are not displayed in the table for brevity.



\begin{figure}[hb]
    \centering
    \includegraphics[scale=\scaleplot]{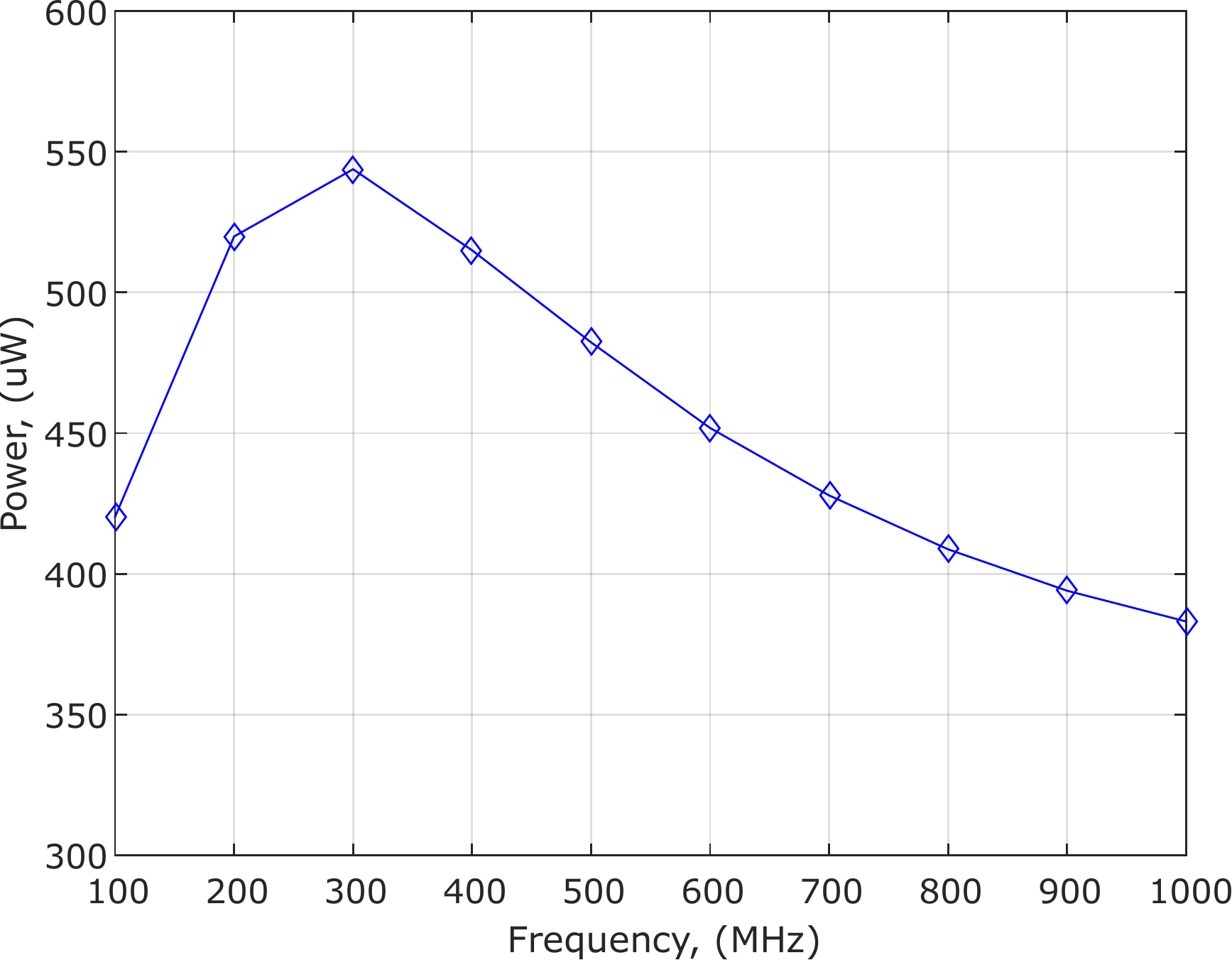}
    \caption{Average power consumption vs input frequency.
}
    \label{fig:pwr_freq}
\end{figure}

The Fig.~\ref{fig:pwr_freq} shows the power consumption of the designed perceptron for different frequencies. The range of the power may vary within several orders of magnitude depending on the parameters of the perceptron such as sizes of the output resistor and capacitor.



\section{Conclusion and Discussions}
\label{sec:conclusions}

We proposed the first mixed-signal (analogue\slash digital) perceptron design using the principle of PWM. Central to our design are a number of parallel inverters that suitable transcode the input-weight pairs from spatial domain to temporal domain. Since PWM-based inverters are typically agnostic to amplitude and frequency variations, the perceptron shows a high degree of power elasticity and robustness under these variations. 

Another advantage of the proposed design is its simplicity. While the conventional implementations of the perceptron require complex logic to perform the multiplication and addition, the proposed approach uses only one gate for per bit for every input. Thus, for the $3 \times 3$ weighted adder we used only 54 transistors. This significantly reduces the logic utilization and, thereafter, the power consumption of the entire device.

Machine learning is finding more applications at the micro-edge, where power variation from energy harvesters is becoming commonplace. We believe the proposed perceptron will find practical implementations in these applications as it is highly robust to these variations. This design would nicely complement a power-elastic PWM signal generator based on a self-timed loadable modulo N counter presented in~\cite{benafa-async2018}.

\bibliographystyle{IEEEtran}
\bibliography{refs}

\begin{thebibliography}{10}
\providecommand{\url}[1]{#1}
\csname url@samestyle\endcsname
\providecommand{\newblock}{\relax}
\providecommand{\bibinfo}[2]{#2}
\providecommand{\BIBentrySTDinterwordspacing}{\spaceskip=0pt\relax}
\providecommand{\BIBentryALTinterwordstretchfactor}{4}
\providecommand{\BIBentryALTinterwordspacing}{\spaceskip=\fontdimen2\font plus
\BIBentryALTinterwordstretchfactor\fontdimen3\font minus
  \fontdimen4\font\relax}
\providecommand{\BIBforeignlanguage}[2]{{%
\expandafter\ifx\csname l@#1\endcsname\relax
\typeout{** WARNING: IEEEtran.bst: No hyphenation pattern has been}%
\typeout{** loaded for the language `#1'. Using the pattern for}%
\typeout{** the default language instead.}%
\else
\language=\csname l@#1\endcsname
\fi
#2}}
\providecommand{\BIBdecl}{\relax}
\BIBdecl

\bibitem{Hagan:1997:NND:249049}
M.~T. Hagan, H.~B. Demuth, and M.~Beale, \emph{Neural Network Design}.\hskip
  1em plus 0.5em minus 0.4em\relax Boston, MA, USA: PWS Publishing Co., 1996.

\bibitem{366063}
E.~Wilson and D.~W. Tufts, ``Multilayer perceptron design algorithm,'' in
  \emph{Proceedings of IEEE Workshop on Neural Networks for Signal Processing},
  1994, pp. 61--68.

\bibitem{doi:10.1111/j.1467-8667.1989.tb00026.x}
\BIBentryALTinterwordspacing
H.~Adeli and C.~Yeh, ``Perceptron learning in engineering design,''
  \emph{Computer-Aided Civil and Infrastructure Engineering}, vol.~4, no.~4,
  pp. 247--256. [Online]. Available:
  \url{https://onlinelibrary.wiley.com/doi/abs/10.1111/j.1467-8667.1989.tb00026.x}
\BIBentrySTDinterwordspacing

\bibitem{HUNG19913}
\BIBentryALTinterwordspacing
S.~Hung and H.~Adeli, ``A model of perceptron learning with a hidden layer for
  engineering design,'' \emph{Neurocomputing}, vol.~3, no.~1, pp. 3 -- 14,
  1991. [Online]. Available:
  \url{http://www.sciencedirect.com/science/article/pii/0925231291900165}
\BIBentrySTDinterwordspacing

\bibitem{330393}
B.~Jeong and Y.~H. Lee, ``Design of weighted order statistic filters using the
  perceptron algorithm,'' \emph{IEEE Transactions on Signal Processing},
  vol.~42, no.~11, pp. 3264--3269, 1994.

\bibitem{1243906}
W.~Qinruo, Y.~Bo, X.~Yun, and L.~Bingru, ``The hardware structure design of
  perceptron with fpga implementation,'' in \emph{SMC'03 Conference
  Proceedings. 2003 IEEE International Conference on Systems, Man and
  Cybernetics. Conference Theme - System Security and Assurance (Cat.
  No.03CH37483)}, vol.~1, 2003, pp. 762--767 vol.1.

\bibitem{8330023}
R.~Shafik, A.~Yakovlev, and S.~Das, ``Real-power computing,'' \emph{IEEE
  Transactions on Computers}, vol.~67, no.~10, pp. 1445--1461, 2018.

\bibitem{benafa-async2018}
O.~Benafa, D.~Sokolov, and A.~Yakovlev, ``Loadable {K}essels counter,'' in
  \emph{Proceedings of ASYNC 2018}, Vienna, May 2018.

\bibitem{7551398}
R.~LiKamWa, Y.~Hou, Y.~Gao, M.~Polansky, and L.~Zhong, ``Redeye: Analog convnet
  image sensor architecture for continuous mobile vision,'' in \emph{2016
  ACM/IEEE 43rd Annual International Symposium on Computer Architecture
  (ISCA)}, 2016, pp. 255--266.

\bibitem{isscc_2016_chen_eyeriss}
{Chen, Yu-Hsin and Krishna, Tushar and Emer, Joel and Sze, Vivienne},
  ``{Eyeriss: An Energy-Efficient Reconfigurable Accelerator for Deep
  Convolutional Neural Networks},'' in \emph{{IEEE International Solid-State
  Circuits Conference, ISSCC 2016, Digest of Technical Papers}}, {2016}, pp.
  {262--263}.

\end{thebibliography}
\end{document}